\documentclass[11pt]{article}
\usepackage{moriond}
\usepackage{graphicx}
\usepackage{epsfig}
\usepackage{epstopdf}

\def\MyPhoto{\pdfimage width 35mm {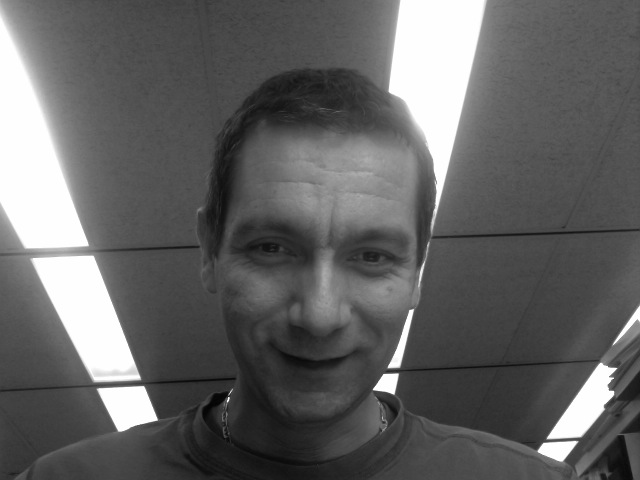}}

\def\be{\begin{equation}}
\def\ee{\end{equation}}
\def\bea{\begin{eqnarray}}
\def\eea{\end{eqnarray}}

\begin{document}
\vspace*{4cm}
\title{MiniBooNE Oscillation Results}

\author{Zelimir Djurcic (for the MiniBooNE Collaboration)}

\address{Department of Physics, Columbia University, New York, NY 10027, USA}

\maketitle\abstracts{These proceedings summarize the MiniBooNE  $\nu_{\mu} \rightarrow \nu_e$ results, describe the first $\bar{\nu}_{\mu} \rightarrow \bar{\nu}_e$ result, 
and current analysis effort with the NuMI neutrinos detected in the MiniBooNE detector.}

\section{Introduction}
Motivated by the LSND observation of an excess of observed $\bar{\nu}_e$ events above Monte Carlo prediction in a $\bar{\nu}_{\mu}$ beam \cite{lsnd}, 
the MiniBooNE experiment was designed to test the neutrino oscillation interpretation of the LSND signal in both neutrino
and anti-neutrino modes. MiniBooNE has approximately the same $\mathrm{L/E_{\nu}}$ as LSND, where L is the neutrino travel
distance and E$_{\nu}$ is the neutrino energy.
However, the MiniBooNE experiment is constructed with an order of magnitude higher
baseline and energy. Due to the higher energy and different event signature, MiniBooNE systematic errors are completely different from LSND errors.

\section{MiniBooNE Neutrino Results}
\begin{figure*}[htb]
\centering
\includegraphics[width=65mm,height=39mm]{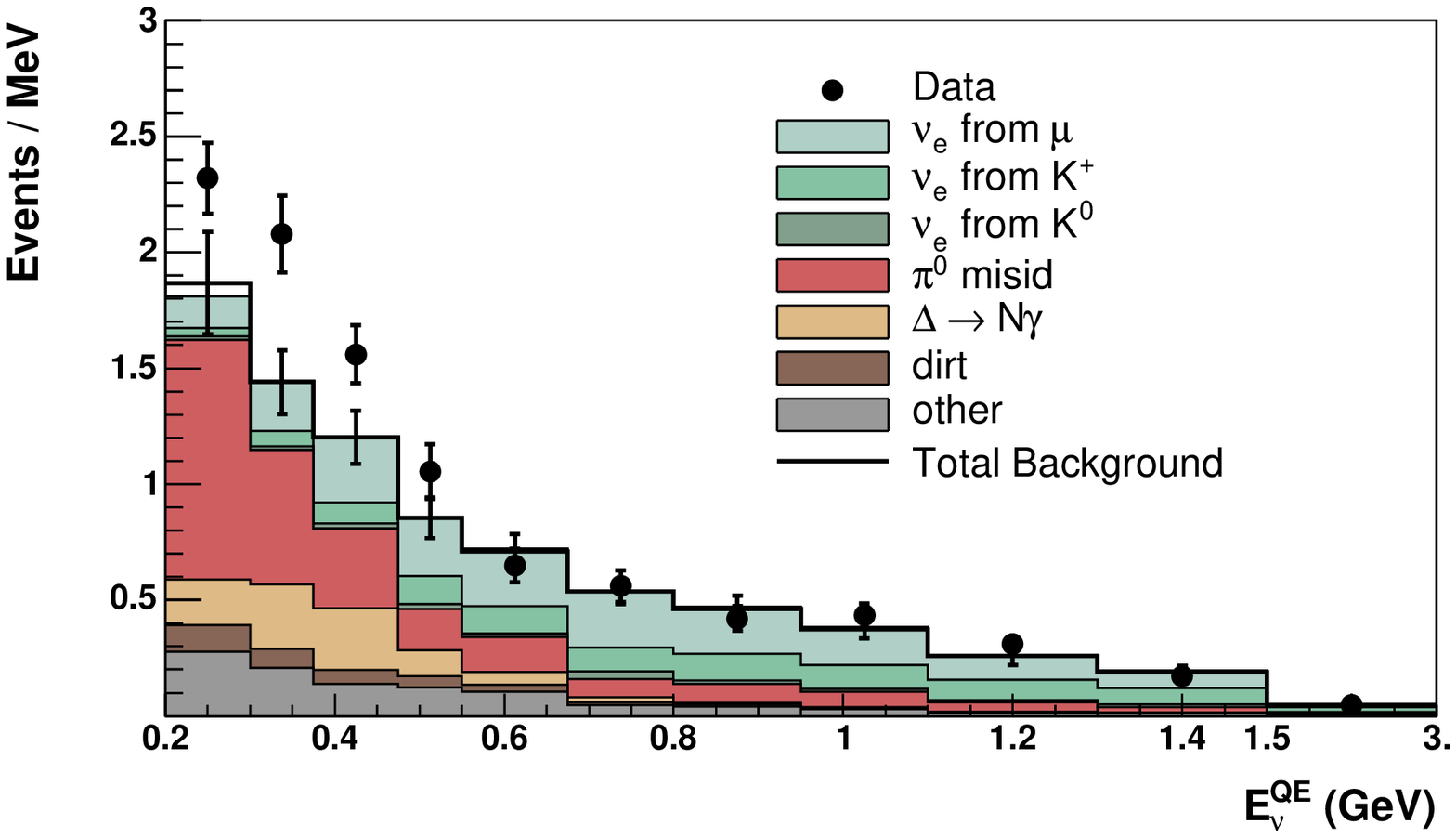}\includegraphics[width=65mm,height=39mm]{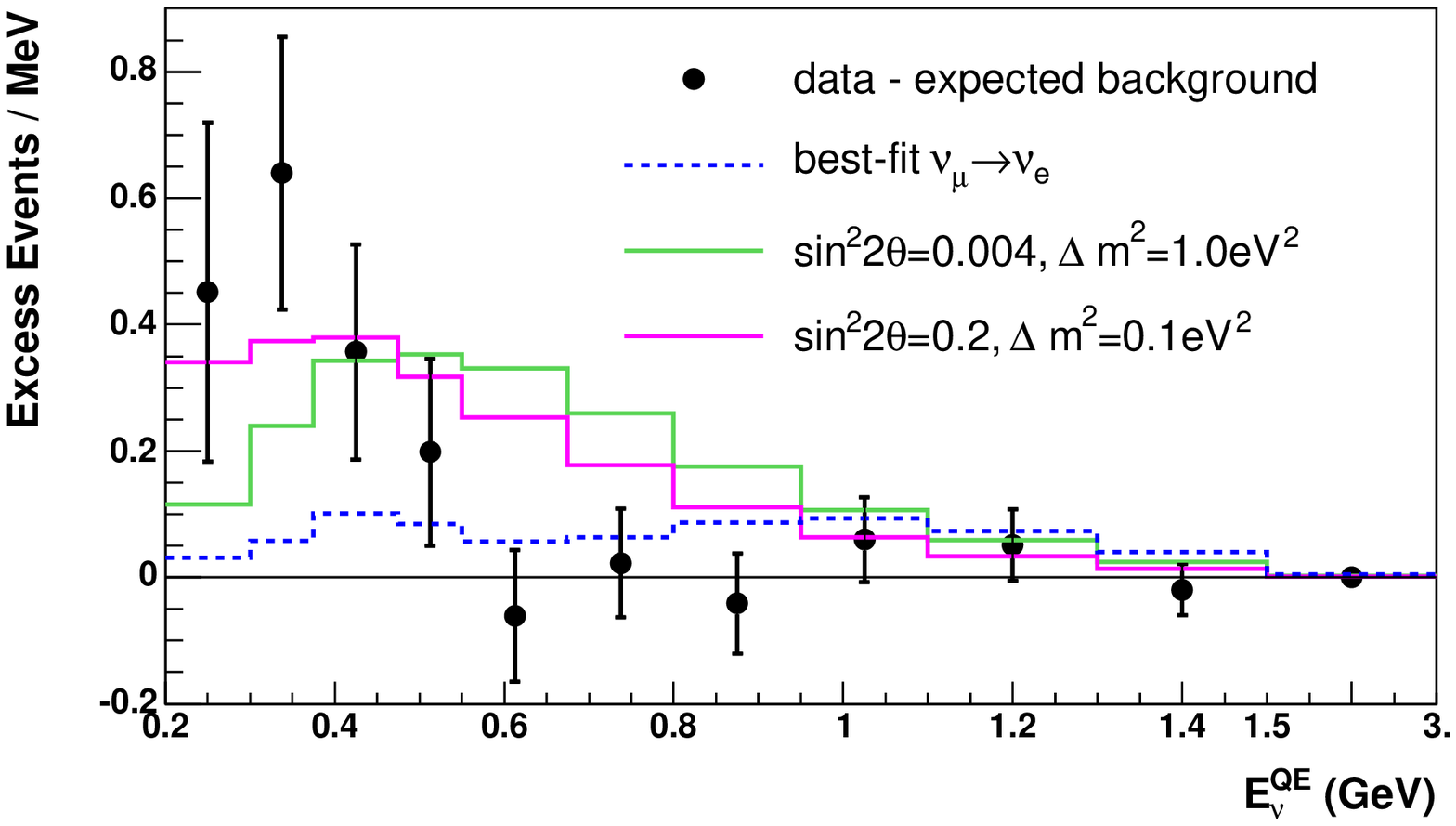}
\caption{ Left: Reconstructed $E_\nu$ distribution of $\nu_e$ CCQE candidates in MiniBooNE neutrino running.  
The data is shown as the points with statistical error. The background prediction is shown as the histogram with systematic errors.
Right: The difference between the data and predicted backgrounds as a function of reconstructed neutrino energy. The error bars include both
statistical and systematic components. Also shown in the figure are expectations from the 
best oscillation fit and from neutrino oscillation parameters in the LSND allowed region.}
\label{fig:MBnew_figure1}
\end{figure*}

The MiniBooNE collaboration has performed a search for $\nu_\mu \rightarrow \nu_e$ oscillations with $6.486 \times 10^{20}$ protons on target (POT), 
the results of which showed no evidence of an excess of $\nu_e$ events for neutrino energies above 475 MeV \cite{MB_osc,BNB_nu_lowE}. 
Fig.~\ref{fig:MBnew_figure1} shows reconstructed $E_{\nu}$ distribution of $\nu_{e}$ CCQE candidates (left).
The right panel of Fig.~\ref{fig:MBnew_figure1} shows the difference between the data and predicted backgrounds as a function of reconstructed neutrino energy. 
Table~\ref{three_BNB_nu_EnuQE_bins} shows observed data and predicted background event numbers in three $E_{\nu}$
bins. The total background is broken down into intrinsic $\nu_{e}$ and $\nu_{\mu}$ induced components.
The $\nu_{\mu}$ induced background is further broken down into its separate components.
\begin{table}[htb!!!]
\centering
\scriptsize
\begin{tabular}[c]{| llll |}\hline
$E_{\nu}$[GeV]                                    & 0.2-0.3              & 0.3-0.475             & 0.475-1.25           \\\hline
Total Bkgd                                            & 186.8$\pm$26 & 228.3$\pm$24.5 & 385.9$\pm$35.7 \\
$\nu_{e}$ induced                              & 18.8                   & 61.7                       & 248.9                    \\
$\nu_{\mu}$ induced                          & 168                   & 166.6                     & 137                        \\\hline
NC $\pi^{0}$                                         & 103.5               & 77.8                       & 71.2                        \\
NC $\Delta\rightarrow N \gamma$  & 19.5                  & 47.5                       & 19.4                       \\
Dirt                                                         & 11.5                  & 12.3                       &  11.5                     \\
Other                                                     & 33.5                  & 29                           & 34.9                      \\\hline
Data                                                      & 232                   & 312                        &  408                       \\\hline
Data-MC                                              & 45.2$\pm$26  & 83.7$\pm$24.5    &  22.1$\pm$35.7  \\\hline
Significance                                        & 1.7$\sigma$    & 3.4$\sigma$         & 0.6$\sigma$       \\ \hline
\end{tabular}
\caption{Observed data and predicted background event numbers in three $E_{\nu}$
bins. In the top rows, the total background is separated into the intrinsic $\nu_{e}$ and $\nu_{\mu}$ induced components.
In the middle rows, the $\nu_{\mu}$ induced background is further broken down into its separate components.\label{three_BNB_nu_EnuQE_bins}}
\end{table}
Despite having observed no evidence for oscillations above 475 MeV, the MiniBooNE $\nu_{\mu}\rightarrow\nu_e$ search observed a sizable excess (128.8$\pm$43.4 events) at low energy, 
between 200-475 MeV \cite{BNB_nu_lowE}. 
Although the excess is incompatible with LSND-type oscillations, several hypotheses, including sterile neutrino oscillations with CP violation \cite{sorel}, anomaly-mediated neutrino-photon coupling \cite{hhh}, and others \cite{weiler,goldman,nelson,hollenberg}, have been proposed that provide a possible explanation for the excess itself. In some cases, these theories offer the possibility of reconciling the MiniBooNE $\nu_e$ excess with the LSND $\bar{\nu}_e$ excess. 
Assuming no CPT or CP violation, the lack of the excess at higher energies allowed 
MiniBooNE to exclude the LSND excess interpreted as two-neutrino oscillations at $\Delta m^2 \sim$ 0.1-100 $\mathrm{eV}^{2}$ at 98\% CL~\cite{heather}.

\section{MiniBooNE Anti-neutrino Results}

\begin{figure*}[htb]
\centering
\includegraphics[width=65mm,height=70mm]{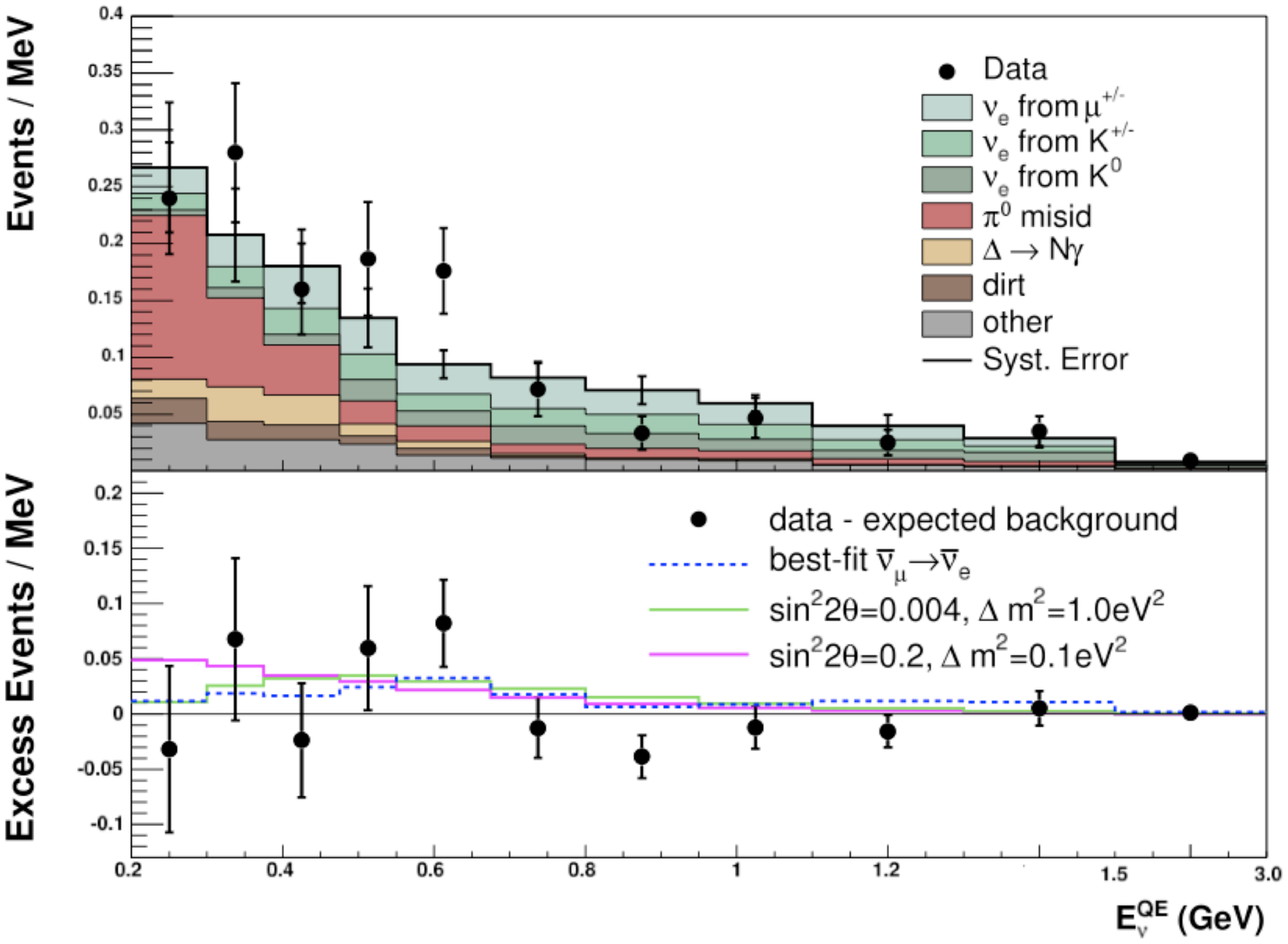}\includegraphics[width=65mm,height=60mm]{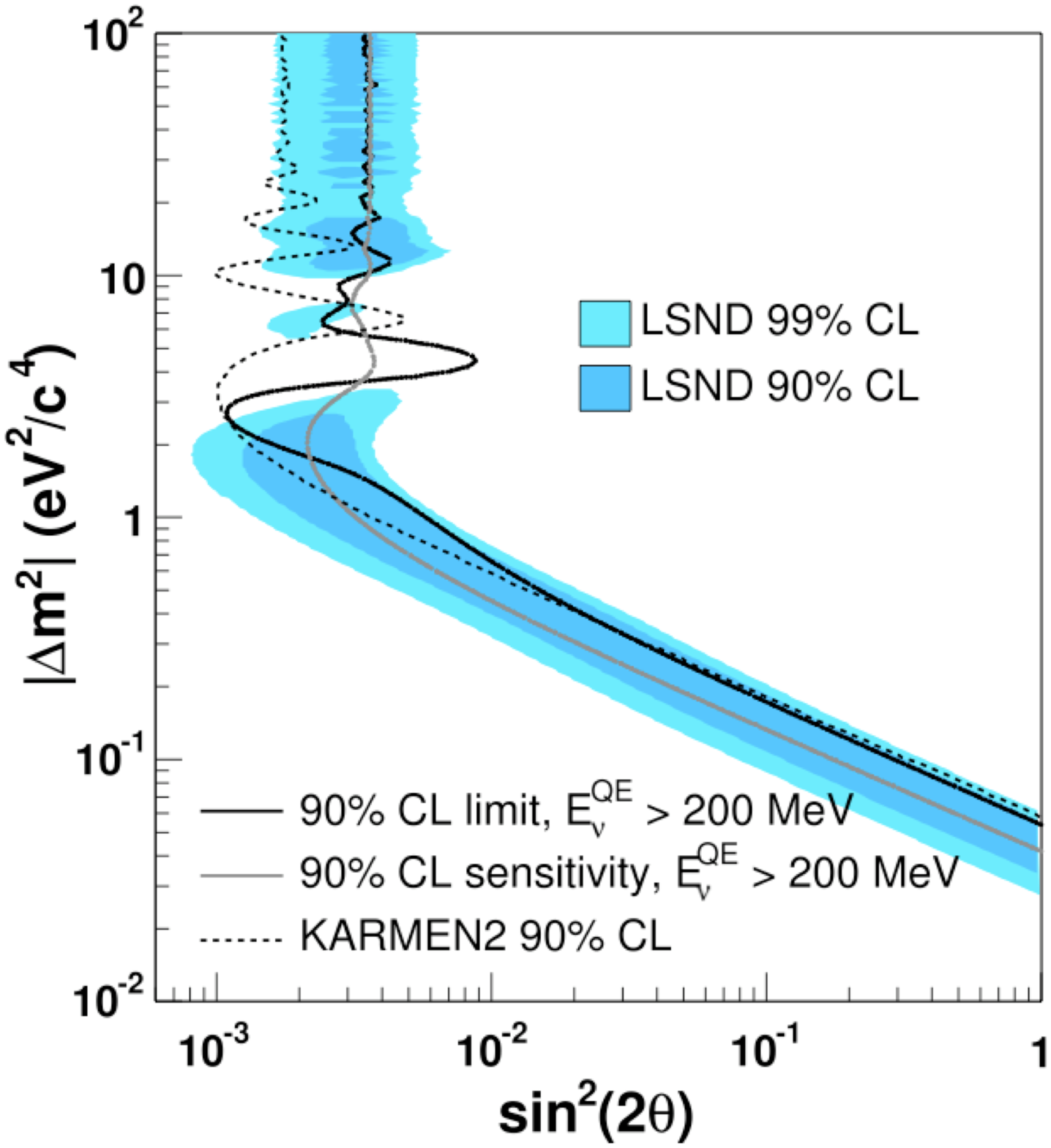}
\caption{ Left Top: Reconstructed $E_\nu$ distribution of $\nu_e$ CCQE candidates in MiniBooNE anti-neutrino running.  
Left Bottom: The difference between the data and predicted backgrounds as a function of reconstructed neutrino energy. The error bars include both
statistical and systematic components. Also shown in the figure are expectations from the 
best oscillation fit and from neutrino oscillation parameters in the LSND allowed region. 
Right: Expected sensitivity and the limit to $\bar{\nu}_{\mu}\rightarrow\bar{\nu}_e$ oscillations from fit to the energy distribution, $E_{\nu}$.
}
\label{fig:MBnuebar_figure1}
\end{figure*}

In December 2008, the MiniBooNE Collaboration also reported initial results from a search for 
$\bar{\nu}_{\mu}\rightarrow\bar{\nu}_e$ oscillations~\cite{MB_anti_nu},
using a data sample corresponding to $3.386 \times 10^{20}$ POT. 
The data are consistent with background prediction across 
the full range of reconstructed
neutrino energy, $200 < E_{\nu} < 3000$ MeV: 144 electron-like
events have been observed in this energy range, compared to an expectation 
of $139.2 \pm 17.6$ events. Fig.~\ref{fig:MBnuebar_figure1} (left) shows reconstructed $E_{\nu}$ distribution of $\nu_{e}$ CCQE candidates.
Table~\ref{two_BNB_antinu_EnuQE_bins} shows observed data and predicted background event numbers in two $E_{\nu}$ bins.  Fig.~\ref{fig:MBnuebar_figure1} (right) shows the expected sensitivity 
and the limit to $\bar{\nu}_{\mu}\rightarrow\bar{\nu}_e$ oscillations from fit to the energy distribution, $E_{\nu}$.
\begin{table}[htb!!!]
\centering
\scriptsize
\begin{tabular}[c]{| lll |}\hline
$E_{\nu}$[GeV]                                                 & 0.2-0.475             & 0.475-1.25           \\\hline
Total Bkgd                                                          & 61.5$\pm$11.7 & 57.8$\pm$10.0 \\
$\stackrel{\small{(-)}}{\nu_e}$ induced         & 17.7                    & 43.1                   \\
$\stackrel{\small{(-)}}{\nu_\mu}$ induced    & 42.6                    & 14.5                        \\\hline
NC $\pi^{0}$                                                      & 24.6                 & 7.2                        \\
NC $\Delta\rightarrow N \gamma$               & 6.6                  & 2.0                       \\
Dirt                                                                      & 4.7                  &  1.9                     \\
Other                                                                  & 6.7                  & 3.4                      \\\hline
Data                                                                   & 61                   &  61                       \\\hline
Data-MC                                                           & -0.5$\pm$11.7  &  3.2$\pm$10.0  \\\hline
Significance                                                      & -0.04$\sigma$    & 0.3$\sigma$       \\ \hline
\end{tabular}
\caption{Observed data and predicted background event numbers in two $E_{\nu}$
bins. In the top rows, the total background is separated into the intrinsic $\stackrel{\small{(-)}}{\nu_e}$ 
and $\stackrel{\small{(-)}}{\nu_\mu}$ induced components.
In the middle rows, the $\stackrel{\small{(-)}}{\nu_\mu}$ induced background is further broken down into its separate components.\label{two_BNB_antinu_EnuQE_bins}}
\end{table}
No significant excess of events 
has been observed, both at low energy, 200-475 MeV, and at high energy, 
475-1250 MeV, although the data are inconclusive with respect to antineutrino 
oscillations at the LSND level.
The 475-675 MeV region shows the data fluctuation of a 2.8 $\sigma$ above the predicted background, 
resulting with the MiniBooNE $\bar{\nu}_{\mu}\rightarrow\bar{\nu}_e$ oscillation limit being worse than
the sensitivity at lower $\Delta m^2$.
These preliminary results, with the excess observed in neutrino mode and the lack of excess in anti-neutrino mode, 
are surprising and suggest that there may be an unexpected difference between neutrino and anti-neutrino properties.\\
It is possible to perform a first comparison of neutrino and anti-neutrino results in the low energy region, 200-475 MeV, and ask
how consistent are the anti-neutrino and neutrino excesses under different assumptions (models).
For example, it may be speculated that the excess of events in the neutrino mode comes from an interaction resulting in an event rate proportional
to the number of protons on the MiniBooNE target (POT). The number of protons on target in neutrino and antineutrino mode is 6.486$\times$10$^{20}$,
and 3.386$\times$10$^{20}$, respectively.
If the excess of 128.8$\pm$43.4 events observed in the neutrino mode is scalable with number of protons, then one would expect
about $128.8\times(3.386\times10^{20} /  6.486\times10^{20})$ = 67 excess events in anti-neutrino mode.
However, such excess was not observed.
Statisticaly, the simplest comparison is in the form a two bin ( one bin for $\nu$, another one for $\bar{\nu}$ data and Monte Carlo) $\chi^2$ test for each assumption
with corresponding errors being statistical only, and with systematic errors fully correlated or uncorrelated.
Table~\ref{nu_nubar_comparison} gives a $\chi^2$ probability assuming one degree of freedom for testing
the following hypotheses as an explanation of the low energy events in neutrino and anti-neutrino modes: the excess 
comes from an interaction resulting in the event rate proportional
to the number of protons on the MiniBooNE target (POT scaled), from a neutral current process with same cross-section for
neutrino and anti-neutrino interaction (Same $\nu$,$\bar{\nu}$ NC), from underestimated neutral current $\pi^{0}$ background (NC $\pi^{0}$ scaled),
from underestimated total background (Bkgd scaled), from an underestimate of kaon flux in at low energies (Low-E Kaons), or 
from  underestimated number of neutrino events in both neutrino and anti-neutrino runs ($\nu$ scaled).
\begin{table}[htb!!!]
\centering
\scriptsize
\begin{tabular}[c]{| lllll |} \hline
Hypothesis	                        & Stat Only  & Cor. Syst	& Uncor. Syst	& Number $\nu$ Expec. \\ \hline
POT scaled	                        &        0.0\%	   &	0.0\%	&       1.8\%      &       67.5\\
Same $\nu$,$\bar{\nu}$ NC&          0.1\%    &    0.1\%        &       6.7\%      &       37.2\\
NC $\pi^{0}$ scaled	               &         3.6\%	   &	6.4\%	&	21.5\%    &       19.4\\
Bkgd scaled	                        &  2.7\%	   &	4.7\%	&	19.2\%	&       20.9\\
CC scaled                                 & 2.9\%	   &	5.2\%	&       19.9\%      &       20.4\\
Low-E Kaons                           & 0.1\%             &    0.1\%       &          5.9\%      &       39.7\\
$\nu$ scaled                             & 38.4               &   51.4\%      &      58.0\%       &        6.7\\ \hline
\end{tabular}
\caption{The $\chi^2$ probability assuming one degree of freedom for testing various hypotheses (described in the text)
as an explanation of the low energy events in neutrino and anti-neutrino modes.\label{nu_nubar_comparison}}
\end{table}
From the simple comparison given in Table~\ref{nu_nubar_comparison} one can see that the hypothesis with the highest probability is the one
where the low energy excess originates from only neutrinos in the beam.
However, more rigorous analysis of the low energy excess, currently underway, is needed to make a strong statement on the nature of the low energy excess.
As of June 2009, the MiniBooNE experiment has collected a total of $5.0 \times 10^{20}$ POT, and has been approved for further running to collect a total of $10.0 \times 10^{20}$ POT 
in anti-neutrino mode. 

\section{MiniBooNE NuMI Results}

\begin{figure*}[htb]
\centering
\includegraphics[width=120mm,height=100mm]{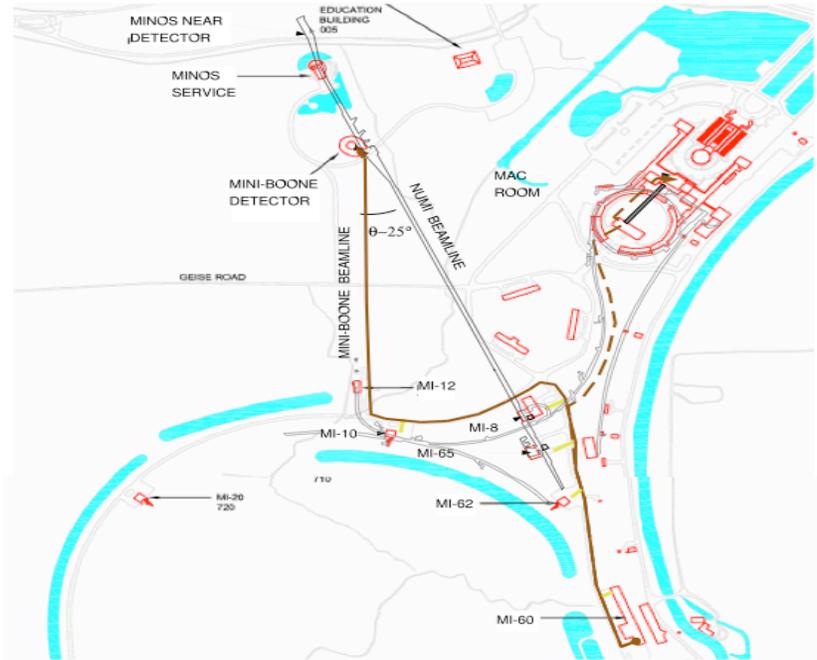}
\caption{Fermi Nation Accelerator Laboratory is currently running two beam lines that produce neutrinos. 
The BNB produces neutrinos
used in the MiniBooNE experiment. 
The NuMI Beam is emitting neutrinos intended for use in the MINOS experiment.} 
\label{fig:beamlines}
\end{figure*}

\begin{figure*}[htb!!!]
\centering
\includegraphics[width=80mm,height=50mm]{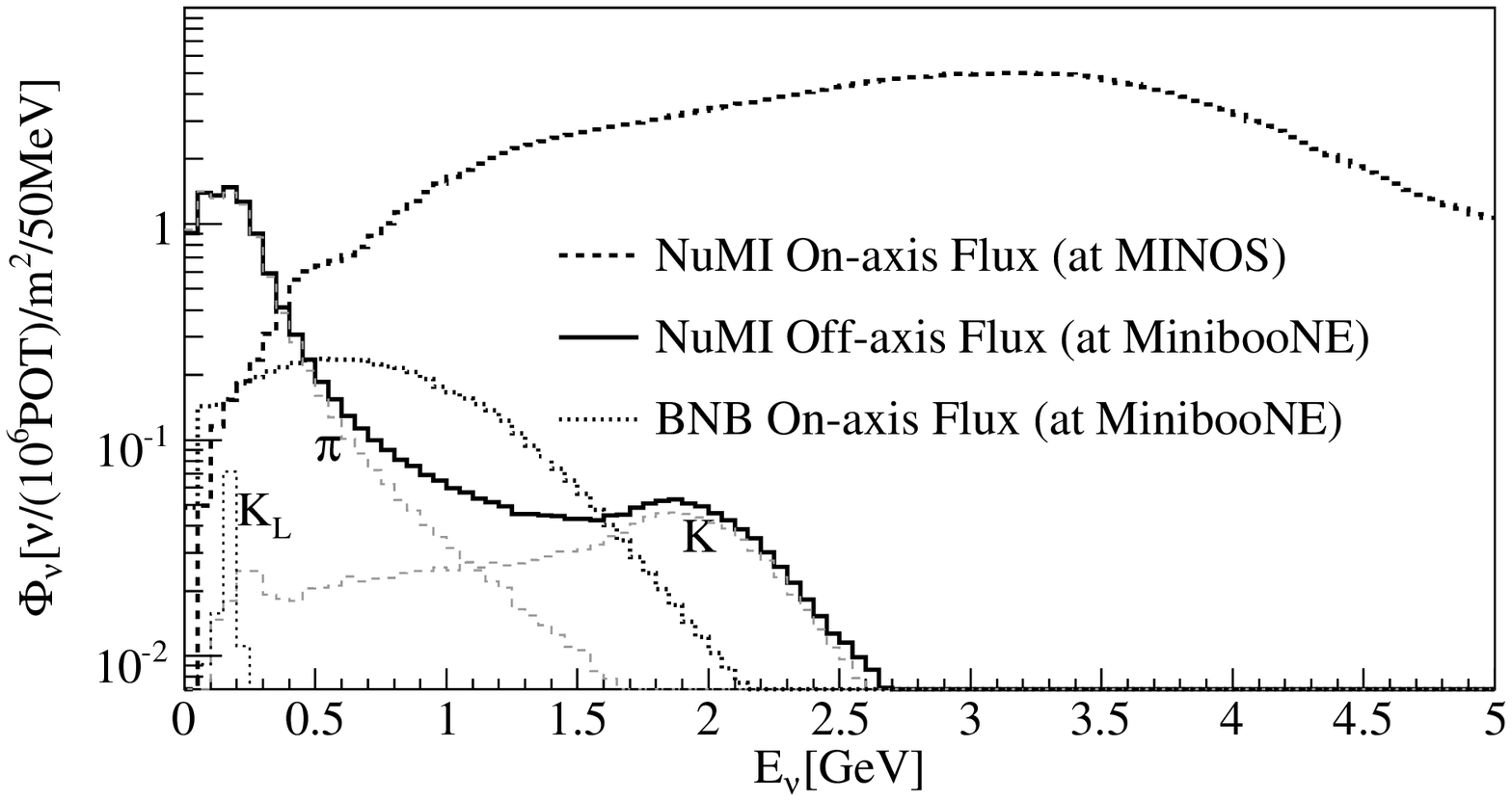}\includegraphics[width=80mm,height=50mm]{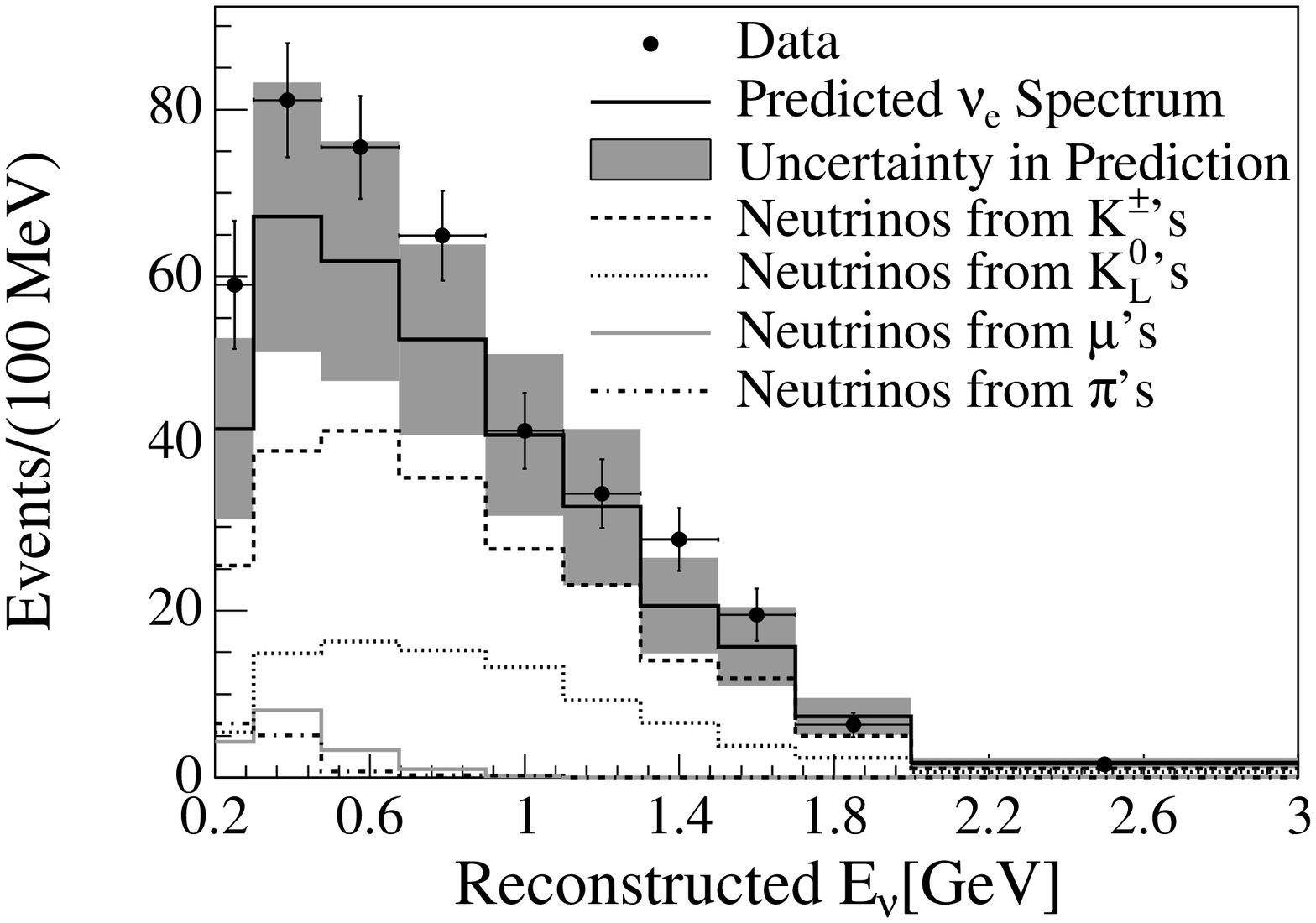}
\caption{Left: Comparison of the predicted NuMI off-axis, NuMI on-axis, and MiniBooNE fluxes including all neutrino species.
The off-axis flux is separated into contributions from charged $\pi$ and $K$ parents. 
Right: Reconstructed $E_{\nu}$ distribution of the NuMI off-axis $\nu_{e}$ CCQE candidate events in MiniBooNE. 
The prediction is separated into contributions from neutrino parents.
The band indicates the total systematic uncertainty associated with the MC prediction. 
Kaon parents contribute 93\% of the events in this sample.}
\label{fig:flux_comparison}
\end{figure*}

An additional data sample measured by the MiniBooNE detector comes from neutrinos
produced in the NuMI (Neutrinos from Main Injector) beam line.
Fermi National Accelerator Laboratory has two beam lines that produce neutrinos: the Booster Neutrino Beam (BNB) and the NuMI beam line,
as shown in Fig.~\ref{fig:beamlines}. The BNB beam is designed for use by the MiniBooNE experiment. 
The NuMI beam produces neutrinos for the MINOS
experiment. However, the MiniBooNE detector observes neutrinos from the NuMI beamline, 
at an off-axis angle of 6.3 degrees.
The NuMI neutrino flux at the MiniBooNE detector is shown in Fig.~\ref{fig:flux_comparison}.
Samples of charged current quasi-elastic (CCQE) $\nu_{\mu}$ and $\nu_e$ interactions were analyzed. 
The high rate and simple topology of  $\nu_{\mu}$ CCQE events
provided a useful sample for understanding
the $\nu_{\mu}$ spectrum and verifying the MC prediction for 
$\nu_{e}$ production. 
The $\nu_e$ CCQE sample energy distribution is shown in Fig.~\ref{fig:flux_comparison}.
These results are described elsewhere~\cite{djurcic_ichep08,first_NuMI} and show that reliable predictions for an off-axis beam 
can be made.

After the demonstration of the off-axis concept, useful in limiting backgrounds in searches for the oscillation transition $\nu_{\mu} \rightarrow \nu_e$, the analysis
is directed toward examing the low energy region and searching for oscillation.
In this way it  complements the analysis done at MiniBooNE using the BNB neutrino and anti-neutrino BNB, but with different systematics.
The phenomenological interpretations of the MiniBooNE results, already mentioned, 
as well as the excess of events observed in neutrino mode and the lack of excess in anti-neutrino mode, 
have provided additional motivation for a neutrino appearance search at MiniBooNE using neutrinos from the NuMI beamline.
It is important to note that the NuMI $\nu_{e}$ CCQE sample has a very different composition
when compared to the BNB neutrino $\nu_{e}$ CCQE sample. The BNB $\nu_{e}$ sample
originates mostly from decays of pions produced in the target, and contains large fraction
of $\nu_{\mu}$ mis-identified events. The NuMI $\nu_{e}$ CCQE sample is produced
mostly from the decay of kaons, and contains a dominant fraction of intrinsic $\nu_{e}$ events.

The analysis will be  performed by forming a correlation between the large statistics $\nu_{\mu}$ CCQE sample 
and $\nu_{e}$ CCQE, and by tuning the prediction to the data simultaneously. 
Considering various sources of systematic uncertainty, a covariance matrix in bins of $E_{\nu}$ is constructed,
which includes correlations between $\nu_e$ CCQE (oscillation signal and background) and $\nu_{\mu}$ CCQE samples. 
This covariance matrix is used in the $\chi^2$ calculation of the oscillation fit.
The result is that the prediction is being constrained, i.e. tuned to the data, 
and common systematic components in $\nu_e$ and $\nu_{\mu}$ CCQE samples  cancel.
The cancellation results from the fact that the majority of the events in both 
$\nu_e$ and $\nu_{\mu}$ CCQE samples originate from pure charged current interaction
of neutrinos sharing same parent mesons, effectively sharing same cross-section and beam 
systematic components.
This is a method equivalent to forming a ratio between 
near and far detectors in two-detector experiments where the near detector detects $\nu_{\mu}$ CCQE events, 
while the far detector samples $\nu_{e}$ CCQE events.
This analysis is in a preliminary stage and is expected to be completed in the near future.

\section{Conclusion}

MiniBooNE observed an unexplained excess of electron-like events in the low energy region
in neutrino mode.
However, no excess of such events is observed so far at low energies in anti-neutrino mode.
MiniBooNE was approved for additional running in anti-neutrino mode, to collect a total 
of $10 \times 10^{20}$ protons on target.
With this additional data taking, which should continue through 2011, 
as well as with the NuMI neutrinos measured by MiniBooNE
the collaboration will be in a position to determine whether there is an 
anomalous difference between neutrino and anti-neutroino properties.

\section*{Acknowledgments}
I would like to acknowledge the support of Fermilab, the Department of Energy, and
the National Science Foundation.

\end{document}